# Confinement Induced Polarization effects in Valence and Inner-shell Spectra of Atactic Polystyrene

*Sudeshna Chattopadhyay*[1], *A. Datta*[1], *A. Giglia*[2], *N. Mahne*[2], *S. Nannarone*[2]

[1]Surface Physics Division, Saha Institute of Nuclear Physics, *1/AF,* Bidhannagar*, Kolkata 700 064. India.* [2]TASC-INFM, AREA Science Park, S.S. 14, Km 163.5, I-34012, Basovizza (TS), Italy.

Phone: 91-033-2337-5345, Extn 4211, FAX: 91-033-2337-4637, e-mail: sudeshna.chattopadhyay@saha.ac.in

Vacuum ultraviolet (VUV) transmission spectra show a clear polarization effect in π electronic transition in spin coated atactic polystyrene (*aPS*) films of thickness below $4R_g$, where $R_g$ (~20.4nm) is the radius of gyration of the polymer. This transition associated with pendant benzene rings in polystyrene. The polarization effect clearly indicates pendant benzene ring alignment on a macroscopic scale. Study of core electron (1s) transition through near edge x-ray absorption fine structure (NEXAFS) spectroscopy confirms the ordering and shows that the rings are oriented towards out-of-plane direction with a tilt angle ~63° with the sample plane, which is consistent with the observed in-plane (sample surface) VUV polarization. These results indicate the transition of a common polymer, like polystyrene, inherently disordered in the bulk, to an orientationally ordered phase under a certain degree of confinement.

**Introduction**

Thin films of polymers have numerous technological applications ranging from multicolour photographic printing to paints, adhesive, index-matched optical coatings, photoresists, and low dielectric electronic packaging. In comparison to bulk properties, much less is known about the properties of polymers when they are processed into thin films. Typically, the polymer-substrate and polymer-air interfacial energies or kinetic barriers play important roles in determining the morphology and dynamics in thin films. A detailed understanding of thin film properties, such as composition and morphology, and possible deviations from bulk properties is highly desirable.

With regard to their spectroscopic properties [1-4], spin cast confined (thickness, $d < 4R_g$, where $R_g$ is the unperturbed polystyrene gyration radius) atactic polystyrene (*aPS*) behaves more like ordered system than as the amorphous systems it is known to be. We already know from supramolecular chemistry that noncovalent interactions between molecules or molecular moieties and their consequent self-assembly lead to the formation of a variety of superstructures [5]. Likewise, ordered behaviour in nanoconfined *aPS* can be explained through noncovalent interaction among its pendant benzene ring moieties. How this inherently disordered polymer is configured in its confined state to behave like ordered system attracts much attention, because this can give rise to some new kinds of properties of the confined polymer, such as its transport, thermal, optical, mechanical etc. properties. The presence of ordering of chemical groups/ chromophores in polymer films can be clearly understood from the study of polarization effects in valence and near edge core shell electronic spectra [6, 7].

In this communication we have presented a realization of strong polarization effect in confined ($R < 4R_g$) *aPS* in its absorption spectra in the near ultraviolet to vacuum ultraviolet (UV-VUV) regime, corresponding to π electronic transition, associated with pendant benzene rings. Also the study of near edge x-ray absorption fine structure (NEXAFS) spectroscopy confirms the observed orientational ordering of pendant benzene rings in confined *aPS* and indicates the ring planes are predominantly oriented towards the out-of-sample plane direction with a tilt angle ~63°±2° relative to the sample (film) plane.

Organization of the paper is as follows. In the present section we introduce our research perspective and motivation. In the next section we describe preparation of *aPS* films of different thickness (*d*-values) and the various experimental procedures and data analysis schema used by us. The section of 'Results and Discussions' has been divided into three subsections. In the first, we present the results of polarization effect in VUV absorption spectra indicates the ordering of pendant benzene rings in *aPS* due to confinement, i.e., thinning the film below a certain thickness. In second subsection we show the resuls extracted from NEXAFS, which confirms the above results and indicates the orientational ordering in confined *aPS*. We then discuss the observed results in context of previous studies on polystyrene and other polymeric films and also for other complex liquids. In the final section we draw our conclusions about this new state of nanoconfined *aPS*.

**Experimental Section**

Atactic *PS* (mol. wt. $M \approx 560900$, $R_g = 0.272 M^{1/2} \approx 20.4$ nm) [8] was spin-coated on fused quartz plates and on Si wafers from toluene solutions using a photo resist spin-coater (Headway Inc.) to form films with *d* varying from 40nm ($\approx 2R_g$) to 147nm ($\approx 7R_g$), and with air/film and film/substrate interfacial roughness ~ 0.6nm, as has been described previously [3]. Before starting experiments with spin-cast *aPS* film Fourier Transform Infra red (FTIR) spectroscopy, with 4.0 $cm^{-1}$ resolution, using a Spectrum GX Perkin Elmer Spectrometer was carried out over the films. As shown in Figure 1, all characteristic peaks of polystyrene in this range were present [9]. However, no peak was present in either the 2870 – 2890 $cm^{-1}$ or the 2950 – 2970 $cm^{-1}$ range, which contain, respectively, the (strong) symmetric and asymmetric stretch bands of the methyl group [9], indicative of the presence of toluene contaminant. Absence of any absorption peak in this range thus rules out the presence of toluene in our *aPS* films. Tacticity of the polystyrene sample were determined from transmission FTIR spectroscopy of the bulk samples on KBr substrate, in the 350-3600 $cm^{-1}$ range, not shown here. The polystyrene sample, were found to be purely atactic, i.e., completely amorphous in its bulk state. Contaminants were removed from the substrate by boiling it with 5:1:1 $H_2O$: $H_2O_2$: $NH_4OH$ solution for 10 minutes, followed by rinsing in acetone and ethyl alcohol.

Grazing incidence x-ray reflectivity data of polystyrene films with $R_g \leq d \leq 8R_g$ were collected using an 18 kW rotating anode x-ray generator (Enraf Nonius FR591). The Cu $K_{\alpha 1}$ line at $\lambda = 0.1540562$ nm was selected with a Si(111) monochromator and measurements were carried out in a triple-axis diffractometer (Optix Microcontrole). Specular reflectivity scans, i.e., scans in the plane containing the incident beam and the normal to the sample surface, with incident angle $\alpha_{in}$=scattering angle $\alpha_{sc}$, were performed with $\alpha_{in}$ varying from 0° to 3° in 5 mdeg steps. If $\boldsymbol{q} = \boldsymbol{k_s} - \boldsymbol{k_i}$ is the momentum transfer vector with $\boldsymbol{k_{s(i)}}$ being the scattered (incident) x-ray wave vector, then this geometry makes the components in the sample plane, $q_x = q_y = 0$, and the value of $q_z$ ($= (4\pi/\lambda) \sin \alpha_{in}$), the component normal to the sample surface, vary from 0 to 4.3 nm$^{-1}$. The resolution in $q_z$ is $\simeq 2.9 \times 10^{-3}$ nm$^{-1}$ and the spatial resolution along film depth is given by $2\pi/q_z^{max}$ where $q_z^{max}$ is the maximum value of $q_z$ to which Kiessig fringes are observed. In our experiments this gives a typical value of $\simeq 1.5$ nm for $z$ resolution. We have measured the off-specular background in all the samples. This was found to be about one-third of the specularly scattered intensity at the maximum $q_z$ (0.3 Å$^{-1}$) used. This off-specular background profile was subtracted from the specular profile and this background-subtracted data has been analyzed.

We have analyzed our reflectivity data using the scheme of distorted wave Born approximation (DWBA) [10], which treats the film as a perturbing potential causing the scattering. This method is very sensitive to small density variations, and has been used to detect density variations due to layering in liquids or polymers [11, 12]. Variations of $\rho$ are expected to be small in polystyrene films and, following our previous studies, we have used the DWBA method to extract EDPs for these films and to look for layer formation. In the DWBA scheme we consider the film to be composed of a number of slices of equal thickness with electron density $\rho(z) = \rho_0 + \Delta\rho(z)$, where $\Delta\rho$ varies with slices but is constant in a slice. The specular reflectivity is then given by [12]

$$\Re(q_z) = |\ ir_0(q_z) + (4\pi r_e/q_z)\ [a_t^2(q_z)\Delta\rho(q_z) + a_r^2(q_z)\Delta\rho^*(q_z)]\ |^2 \qquad (6)$$

where $r_0$ is the specular reflectance coefficient of the average film of electron density $\rho_0$, $a_t$ and $a_r$ are the coefficients for transmitted and reflected amplitudes of the average film, and $\Delta\rho(q_z)$ is the Fourier transform of $\Delta\rho(z)$. By selecting an appropriate number of slices (of width always greater than 1.5 nm, the maximum achievable resolution) an appropriate $\rho_0$ for the film, we fit Eqn. (6), after appropriate convolution of the calculated reflectivity for the detector slit, to the data for *pPS* films, keeping $\Delta\rho$'s, air/*PS* and *PS*/*Qtz* gaussian interface widths ($\sigma_{PS}$ and $\sigma_{PS\text{-}Qtz}$) as the fit parameters. The EDP is constructed from the values of $\rho(z)$ obtained from the best fit. The calculated reflectivity profile has also been convoluted by a Gaussian resolution function with a FWHM corresponding to the 400 μm wide detector slit. The veracity of this procedure has been crosschecked by generating the reflectivity profiles from the extracted EDPs using the Parratt scheme [13].

We have carried out VUV spectroscopy of polystyrene films with $d \approx 2R_g$ to $8R_g$. Transmission spectra in the 4-9eV range were collected in 10meV steps at BEAR beamline of ELETTRA synchrotron, with nearly linearly polarized light (the estimated Stokes parameter $S_1 \approx 0.5$), the electric field lying in the film plane [14]. The experimental chamber was maintained at $\sim 10^{-10}$ Torr and all measurements were done at ambient temperature. The VUV transmission spectra were normalized to the incident photon flux. Our focus was on the electronic singlet transition $e_{1g}(\pi) \rightarrow e_{2u}(\pi*)$ involving the pendant benzene rings (shown in inset of Figure 2(a)) of *aPS*, which is centered on 6eV. The transmission spectra were taken at normal incidence ($\theta_M = 90°$) with different $\psi_c$ angle, where $\psi_c$ is the angle between the $Y_S$ and the $Y_L$, shown in Figures 3(a) and 3(b). The impinging light was elliptically polarized with a degree of linear polarization $P = \dfrac{E_H^2 - E_v^2}{E_H^2 + E_v^2} = 0.5$

Where $E_H$ and $E_v$ are the horizontal and vertical components of the electric field of the electromagnetic wave.

The XAS spectra were normalized with the incident photon flux. Intensities were monitored by measuring the current drained by a tungsten grid inserted in the beam path. The incident flux was obtained in a separate measurement by the drain current of a clean Cu sample. The absorption coefficient as a function of incident photon energy was obtained taking the ratio between *aPS* film spectra to that of the clean Cu. The photon energy resolution was 50meV at the C K-edge, and the typical photon flux on the sample was of the order of $10^{10}$ photon/second. The NEXAFS spectra have been measured in total electron yield, drain current mode. The impinging light was elliptically polarized with a degree of linear polarization, $P = 0.7$. In this work the XAS scans were performed in two different geometries, polar scan and precession of normal scan. In the polar scan the measurements were taken by changing the incidence of light with respect to the surface normal. In precession scan the spectra are taken with a constant incidence angle and making the sample normal to precess around the incidence direction (Figures 3(a) and 3(b)) producing a rotation of the polarization ellipse around the incidence direction at the sample surface.

**Results and Discussion**

**1. Pronounced in-plane polarization effect in VUV absorption spectra in π electronic transition**

Polystyrene in its near UV-VUV absorption spectra shows intense peak around 6eV corresponding to strong singlet optical transition arises from particle-hole (molecular exciton) states constructed from the $e_{1g}(\pi)$ and $e_{2u}(\pi^*)$ molecular orbitals in the benzene moiety [15]. In assigning the various polystyrene transitions in Figure 2(a) the one at lowest photon energy is the vibronic (that is orbitally forbidden but vibronically allowed) transition $^1A_{1g} \rightarrow {}^1B_{2u}$ band (~4.8eV). This transition involves combination of purely electronic transition with the vibration of symmetry $e_{2g}$. This absorption is polarized in the plane of the pendant benzene ring. The next transition is that indicated by only a shoulder at 5.79eV corresponds to vibronic $^1A_{1g} \rightarrow {}^1B_{1u}$ transition. This transition involves combination of the purely electronic transition with the vibration of symmetry $e_{2g}$ polarized in the plane of the benzene moiety in polystyrene, as well as combination with the vibrations of symmetry $b_{2g}$, polarized along the perpendicular to the plane of the ring [16]. The only vibrationless electronic transition allowed in the pendant group of the solid *aPS* film is $^1A_{1g} \rightarrow {}^1E_{1u}$, giving rise to the third peak around ~6.3eV, which is the most intense transition [1, 17], shown in the spectra. This transition is polarized in the plane of the pendant benzene ring [16].

Vacuum ultraviolet spectroscopy of *aPS* films of thickness $R < 4R_g$ shows that there is pronounced polarization effect in the assigned electronic transition (as in Figure 2(a)), shown in Figure 4(a). The pure electronic singlet $^1A_{1g} \rightarrow {}^1E_{1u}$, which is polarized in the plane of the pendant benzene ring, and the out-of-plane component of vibronic singlet $^1A_{1g} \rightarrow {}^1B_{1u}$ show opposite polarization (Figure 4(a)) effects (i.e, the intensity of in-ring-plane transition becomes sharper and is clearly resolved when out-of-ring plane transition intensity is lowered) for precession scan with different $\psi_c$ angle with normal incidence ($\theta_M=90°$) of linearly polarized light (i.e., as the linearly polarized electric field is rotated on the film plane). Here, degree of linear polarization, $P=0.5$, at 4 to 9eV photon energy range, as mentioned already in experimental section, i.e., $E_v/E_H= \varepsilon = 0.58$. The strong in-sample-plane polarization effects of in-plane and out-of-plane transition in pendant benzene rings in nanoconfined ($d < 4R_g$) polystyrene film indicate (1) the pendant benzene rings are not a disordered system, rather they show macroscopic orientational order in spin-cast nanoconfined atactic polystyrene film, whereas atactic polystyrene is inherently disordered in its bulk state, and (2) rings must be oriented towards out-of-sample- plane direction. Depending on these observations we have modeled the ordered pendant benzene ring system as follows: normal vector on benzene ring plane or π orbitals of benzene rings are oriented with azimuthal angle $\phi$ and polar angle $\theta$ in sample frame of reference, shown in Figure 3(a). $\theta$ gives the tilt angle of the ring plane with respect to sample plane in other words the pendant benzene rings are oriented with (90°-$\theta$) or α tilt angle relative to the sample surface normal, as shown in Figure 5(b). Electric field components in laboratory frame of reference, $E_{Lx}= 0$, $E_{Ly}= E_H$, and $E_{Lz}= E_v = \varepsilon\, e^{i\delta(\omega)} E_H$, and can be expressed as,

$$\vec{E}_L(\omega) = \vec{E}_H(\omega) e^{i(\omega t - \vec{k}\cdot\vec{r})} \begin{pmatrix} 0 \\ 1 \\ \varepsilon e^{i\delta(\omega)} \end{pmatrix} \qquad (1)$$

Here, δ is the phase between $E_H$ and $E_v$, taken to be 90°. When the sample is rotated about incident direction, $x_L$ axis in the lab frame of reference, with normal incidence (i.e., $\theta_M=90°$), by an angle $\psi_c$, the angle between the $Y_S$ and $Y_L$ axes of samples laboratory reference frames respectively (see Figure 3(b)) then the angular ($\psi_c$) dependence of the resonance intensity of the out-of-ring-plane transition corresponding to vibronic singlet $^1A_{1g} \rightarrow {}^1B_{1u}$ (as discussed in previous section), with transition dipole moment of magnitude $\mu_2$, is given by for single domain

$$I \propto (\mu_2 E_H)^2 \sin^2\theta[(\varepsilon^2\cos^2\psi_c + \sin^2\psi_c)\cos^2\phi + (\cos^2\psi_c + \varepsilon^2\sin^2\psi_c)\sin^2\phi + 2\cos\psi_c\sin\psi_c(\varepsilon^2-1)\cos\phi\sin\phi]$$
$$= A\sin^2\theta[(\varepsilon^2\cos^2\psi_c + \sin^2\psi_c)\cos^2\phi + (\cos^2\psi_c + \varepsilon^2\sin^2\psi_c)\sin^2\phi + 2\cos\psi_c\sin\psi_c(\varepsilon^2-1)\cos\phi\sin\phi]$$
(2)

$A$ is a constant factor. We have plotted the intensity variation of $^1A_{1g} \rightarrow {}^1B_{1u}$ transition for a given film thickness, $d=2R_g$, for different (in-sample-plane) polarization angle $\psi_c$, shown in Figure 4(b), and tried to fit the variation with Eqn.(2) (with $\phi$ as fitting parameter). Fitted result, by Eqn.2, not matched well to the experimental results where as the experimental values are well fitted with consideration of two fold symmetry, where the cross term containing $\cos\phi$ is eliminated upon averaging over domains and then Eqn.(2) becomes

$$I = A\sin^2\theta[(\varepsilon^2\cos^2\psi_c + \sin^2\psi_c)\cos^2\phi + (\cos^2\psi_c + \varepsilon^2\sin^2\psi_c)\sin^2\phi] \quad (3)$$

Ring-plane transition $^1A_{1g} \rightarrow {}^1E_{1u}$ becomes much sharper where as the out-of-ring-plane transition goes down with lowest intensity, completely agreed with our model. This observation is true for all the film thickness of nanoconfined ($d<4R_g$) aPS, where only one characteristic film thickness ($d \approx 2R_g$) is shown in Figure4. In this context it is to be noted here that there is no such type of polarization effect was observed in valence electronic transition in case of thick ($d > 4R_g$) aPS film, shown in Figure 4(c).

**2. Orientation of pendant benzene rings (in nanoconfined *aPS*) with the sample plane from NEXAFS study**

Near-edge X-ray absorption fine structure (NEXAFS) can be used to determine the orientation of pendant benzene rings of nanoconfined polystyrene with respect to the sample plane by probing the dependence of transition intensities into unoccupied molecular orbitals of the pendant molecules on the angle of incidence of linearly polarized light [18, 7].

Figure 2(b) shows the C K edge x-ray absorption spectra (XAS) of nanoconfined polystyrene of film thickness $d = 2R_g$. Spectra were taken at normal incidence with elliptiacally polarized light. The π and σ resonances in the absorption spectra of polystyrene film are assigned [19-22] in Figure 3(a). Overall, the spectrum strongly resembles the spectrum of benzene, which is a large component of the polymer [19], and energy level corresponding to the π* transition in benzene are shown in inset of Figure 2(b).

The tilt angle of the benzene ring planes with respect to sample surface plane was obtained by measuring the absorption coefficient of the nanoconfined polystyrene film at different incidence field configuration with respect to surface of the sample, through the precession and polar scans. Similarly as in the VUV scheme, the angle $\psi_c$ is the angle between the $Y_S$ and the $Y_L$ axes of sample and laboratory reference frames respectively (see, Figure 3b) [23]. A quantitative analysis of NEXAFS for different experimental geometries yields the average tilt angle of the benzene rings with respect to the sample plane [24]. This angle was obtained by first decomposing each absorption coefficient spectrum taken at particular $\psi_c$ by a procedure based on lineshape simulation through a suitable number of Voigt functions and a step function (shown in Figure 3a). Then by fitting the variation of each component with $\psi_c$ using the tilt angle of the benzene ring plane as fitting parameter [23, 7], the average value of this angle was determined. The intensity variation of each component was fitted to the same model as before, where the absorption intensity ($I$) is proportional to $\left|\vec{E}.\vec{p}\right|^2$, where $\vec{p}$ is the vector matrix element of dipole operator between 1s initial state, and π* final orbital state, directed along the final state π orbital dipole vector (i.e., along normal vector to the benzene ring plane), with $\varepsilon = E_v/E_H=0.42$, at 270 to 320eV range, written in the sample frame reference,

The general fitting expression for a single domain is given by:

$$|\vec{E}\cdot\vec{p}|^2 = p^2 E_H^2 \begin{bmatrix} (\varepsilon^2 \cos^2\psi_C + \sin^2\psi_C - 2\varepsilon\cos\psi_C \sin\psi_C \cos\delta)(\sin\theta_M \sin\theta\cos\phi + \cos\theta_M \cos\theta)^2 \\ +(\cos^2\psi_C + \varepsilon^2 \sin^2\psi_C + 2\varepsilon\cos\psi_C \sin\psi_C \cos\delta)\sin^2\theta\sin^2\phi \\ +2\{\cos\psi_C \sin\psi_C(\varepsilon^2-1) + \varepsilon\cos\delta(\cos^2\psi_C - \sin^2\psi_C)\} \\ \{(\sin\theta_M \sin\theta\cos\phi + \cos\theta_M \cos\theta)\sin\theta\sin\phi\} \end{bmatrix}$$
(4)

Here, $\theta_M$ is the complementary angle of the incidence angle and $\delta$ (=90°) the phase between $E_H$ and $E_V$ as mentioned earlier. $\theta$ and $\phi$ are the polar and azimuth angles of the $\vec{p}$ vector in the sample frame of reference. If ring plane is perpendicular to the $\vec{p}$, the tilt angle of the ring plane is $\theta$ w.r.t. sample plane i.e., the tilt angle of the ring plane relative to the sample surface normal is (90°-$\theta$), shown in inset of Figure 5(b). It should be noted here that the Eqn. (4) reduced to Eqn. (2) if we put $\theta_M$=90°, i.e., at normal incidence.

In experiment, we have first taken the energy scan from 270 to 320eV range with $\theta_M$=90°, $\psi_c$=0, and denoted the normalized intensity of $\pi_{1A}$ as $I_0$ (=$I_{\pi 1A}(\theta_M$=90°,$\psi_c$=0)), then for $\theta_M$=20° we have taken the precession scan i.e., the energy scans in above-mentioned range with different $\psi_c$ values, shown in Figure 5(a). Figure 5(a) shows $\pi^*_1$ peak exhibits the strongest polarization dependence which is opposite to that of $\sigma^*$ as expected [7]. The intensity of $\pi^*_1$ transition decreased markedly when $\theta_M$ goes from 90° to 20°, i.e., from normal incidence to grazing incidence at $\psi_c$=0, which indicate clearly that pendant benzene rings are oriented towards out-of-plane direction with respect to the sample surface plane [7]. In Figure 5(b) the $\pi_{1A}$ intensities, normalized by $I_0$, for different $\psi_c$ values, with constant $\theta_M$ (=20°), are plotted. It is observed that the experimental results, in case of inner-shell spectra also, fit well with Equn. (4) when we consider, as in the case of VUV polarization results, two-fold symmetry in the system, i.e., using the equation,

$$\frac{I_{\pi 1A}}{I_0} = \frac{(\varepsilon^2 \cos^2\psi_C + \sin^2\psi_C)(\sin^2\theta_M \sin^2\theta\cos^2\phi + \cos^2\theta_M \cos^2\theta) + (\cos^2\psi_C + \varepsilon^2 \sin^2\psi_C)\sin^2\theta\sin^2\phi}{(\varepsilon^2 \sin^2\theta\cos^2\phi + \sin^2\theta\sin^2\phi)}$$
(5)

for fitting the data. Solid line in Figure 5(b) shows the best fit obtained for the first resonance $\pi_{1A}$ of the nanoconfined polystyrene film ($d<4R_g$) spectra. The fit provides a tilt angle ($\theta$) of the molecular (pendant ring) plane of 63°±2° with respect to the sample plane, shown in inset of Figure 5(b). The same experiments were performed with different film thickness of polystyrene below and above $4R_g$. All the films, of thickness $d < 4R_g$, show the same kind of polarization effect as in case of $d = 2R_g$ film, discussed above. On the other hand, there is no such polarization effect were observed for film thickness $d > 4R_g$ (not shown here). The above results, concerning the polarization effect in valence and inner-shell electron transition spectra, clearly show that due to confinement below a thickness of $4R_g$ there is a macroscopic orientational ordering in the pendant benzene rings of spin cast polystyrene at ambient temperature. The rings are oriented at a tilt angle $\theta = 63°±2°$ with the sample-plane (Figure 5(b), inset).

The transition of atactic polystyrene, which is orientationally disordered in the bulk, to this ordered phase under nanoconfinement, is observed directly through confinement induced polarization effects in valence and inner-shell spectra of the pendant benzene rings, probably for the first time, although syndiotactic-like behaviour had previously been observed in FTIR and dielectric relaxation spectroscopy of atactic polystyrene under confinement [2].

In this connection it should be noted that above the film thickness of $4R_g$, buffing can produce alignment of benzene rings along the direction of the external stress applied and this alignment is most pronounced upto a depth of 1nm from the surface [22]. Our results indicate that under nanoconfinement, ie, film thickness $< 4R_g$, the confinement acts as an external 'field' to cause orientational ordering of the rings and, as we have obtained VUV polarization in the transmission mode, (where total film thickness is probed), this ordering dominates the entire film thickness. It is also interesting to note that the value of tilt angle of the benzene rings obtained in our nanoconfined polystyrene i.e., 27° relative to the sample surface normal, is of the order of the tilt angle (20°) seen through IR-visible Sum Frequency Generation Spectroscopy [25] for benzene rings at the air-polymer interface in polystyrene films. This indicates that the pendant benzene rings in the two cases maybe in similar orientationally ordered phases suggesting a similarity between confinement due to total thinning and that at the surface layer.

Development of orientational order due to confinement has been observed in some organic films. In alkanes such ordering is connected to reduction in gauche defects [26] while for liquid crystalline cellulose films the orientational ordering is found to be correlated to formation of molecular layers under confinement and a minimum film thickness equal to the liquid crystalline unit cell is required for formation of layers as well as for orientational ordering [27]. Our results regarding the development of orientational order under confinement in polystyrene are also found to be correlated to the formation of layers in the film with the periodicity of the molecular gyration radius ($R_g$), which, as we have previously shown, [3, 28] takes place below a film thickness of $4R_g$, exactly when the film also starts to show orientational order.

**Conclusion and Outlook**

We have found from the study of valence and innershell spectra in polystyrene that for film under a thickness of four times the radius of gyration ($4R_g$), where $R_g$=20.4nm, shows clear polarization effects, which indicates orientational ordering of pendant benzene ring. Both the results concerning in-plane polarization effects in VUV spectra and polarization effects in NEXAFS spectra involves the pendant benzene rings in polystyrene, fit well with theoretical model shows that pendant benzene rings in confined atactic polystyrene film are ordered (with two fold symmetry) and make a tilt angle 63°±2° with the sample-plane.

Since thin polymer films are widely used in many applications in modern technology, the miniaturization of components in general requires processing and dealing with polymer films of ever decreasing thickness. Our results shows that under a certain thickness there may be a dramatic change in structure and physical properties of these films and suitability for the specific application requires a thorough knowledge of thin film structure and properties. On the other hand such knowledge may lead to completely new uses of these films as for e.g., polarization dependence of our polystyrene films may open up new optical applications. Again an ordered array of benzene rings as found here can be used as ordered template for supramolecular chemistry [28] and biochemistry.

**Figure 1**: (Color online). Fourier Transform Infrared (FTIR) spectrum (transmittance versus wavenumber) of $2R_g$ ($\approx$ 40 nm) and $8R_g$ ($\approx$ 160 nm) thick atactic polystyrene (*aPS*) films. The strong absorption peaks are assigned as follows: methylene symmetric stretch – 2851 cm$^{-1}$, methylene asymmetric stretch – 2924 cm$^{-1}$ and aromatic hydrocarbon stretch – 3027 cm$^{-1}$.

**Figure 2**: (Color online). Assignment of VUV absorption spectra and Near edge x-ray absorption fine structure (NEXAFS) spectra of polystytrene. (a): Transmission spectra (absorbance versus photon energy in eV) in vacuum ultraviolet (VUV) for *aPS* films on quartz substrate, with film thickness $d$ = 40 nm, for $\theta_M$=90°, $\psi_c$=0°. Assigned transitions presented besides the spectral bands. Circles: data, red line: convolution of individual gaussian fits in green line. Inset: Corresponding transitions and energy levels in case of benzene. (b): Near edge x-ray absorption fine structure (NEXAFS) spectra (absorbance versus photon energy in eV) at the Carbon *K* edge of *aPS* films on Si substrate, with film thickness $d$ = 40 nm for $\theta_M$=90°, $\psi_c$=0°. Assigned transitions presented besides the spectral bands. Circles: data, red line: convolution of individual Voigt fits in blue line. Inset1: zoomed portion of the spectrum to clearly show the $\pi^*_1$ and $\pi^*_2$ transition. Inset2: Corresponding transitions and energy levels in case of benzene.

**Figure 3**: (Color online). Experimental geometry. (a): Coordinate system defining the geometry of the $\pi^*$ vector orbital on the benzene ring plane on the sample surface plane. The orientation of the orbital, i.e., of the vector **O**, is characterized by a polar angle $\theta$ and an azimuthal angle $\phi$. The x-rays are incident in the (X,Z) plane containing the minor electric field vector component $E_v$. The x-ray incidence angle (90°-$\theta_M$) is also the polar angle of $E_v$ and is changed by rotating the sample about the Y axis. The major electric field vector component $E_H$ lies in the surface plane, along the Y axis. (b): experimental geometry of the precession scan. *n* indicates the normal to the surface, $E_H$ and $E_v$ are,

respectively, the horizontal (major) and the vertical components of the elliptically polarized light. The axes of the sample and laboratory frame of references are labeled as S and L respectively.

**Figure 4**: (Color online).(a)-(b): Polarization effect in VUV absorption spectra of *aPS* film of thickness $d \approx 2R_g \approx 40$ nm. (a): Transmission spectra (absorbance versus photon energy in eV) in vacuum ultraviolet (VUV) with different $\psi_c$ values, $\psi_c$=0°, 30°, 60° and 90°, for normal incidence i.e., $\theta_M$=90°. Corresponding $\psi_c$ values for each profile are shown with appropriate colour code. (b): Intensity variation of $^1A_{1g} \rightarrow {}^1B_{1u}$ transition with $\psi_c$ is plotted as filled circles. Intensity is normalized by $I_{90}$, where $I_{90}$ is the intensity of same transition for $\psi_c$ =90°, at normal incidence for same film. The data is fitted with Eqn. (3) (normalized accordingly, refer text) shown by red line. (c): Absorption spectra in VUV with same condition as in (a) for film thickness, $d \approx 8R_g$ (i.e., $d > 4R_g$).

**Figure. 5**: (Color online). Polarization effect in NEXAFS spectra of *aPS* film of thickness $d \approx 2R_g \approx 40$ nm. (a): Near edge x-ray absorption fine structure (NEXAFS) spectra (absorbance versus photon energy in eV) at the Carbon *K* edge of *aPS* films on Si substrate, with film thickness $d$ = 40 nm for $\psi_c$=0°, shown in black and red line for $\theta_M$=90° and $\theta_M$=20° respectively. Inset: zoomed portion of the spectra to clearly show the variation in $\pi^*_1$ transition intensity. (b): For fixed $\theta_M$ (=20° ), focusing on $\pi^*_1$ transition, $\pi_{1A}$ resonance intensity variation with different $\psi_c$ values is plotted as filled circles. Here intensity is normalized with $I_0$, where $I_0$ is the intensity of same transition for $\theta_M$ =90° and $\psi_c$=0°. The data is fitted with Eqn. (5) shown by red line. Best fit was obtained for a tilt angle, $\theta$ = 63°±2°. Inset: pendant benzene ring plane in confined *aPS*, for film thickness $d<4R_g$, oriented relative to the surface normal with $\alpha$ (=90°- $\theta$). For best fit $\alpha$ = 27°.

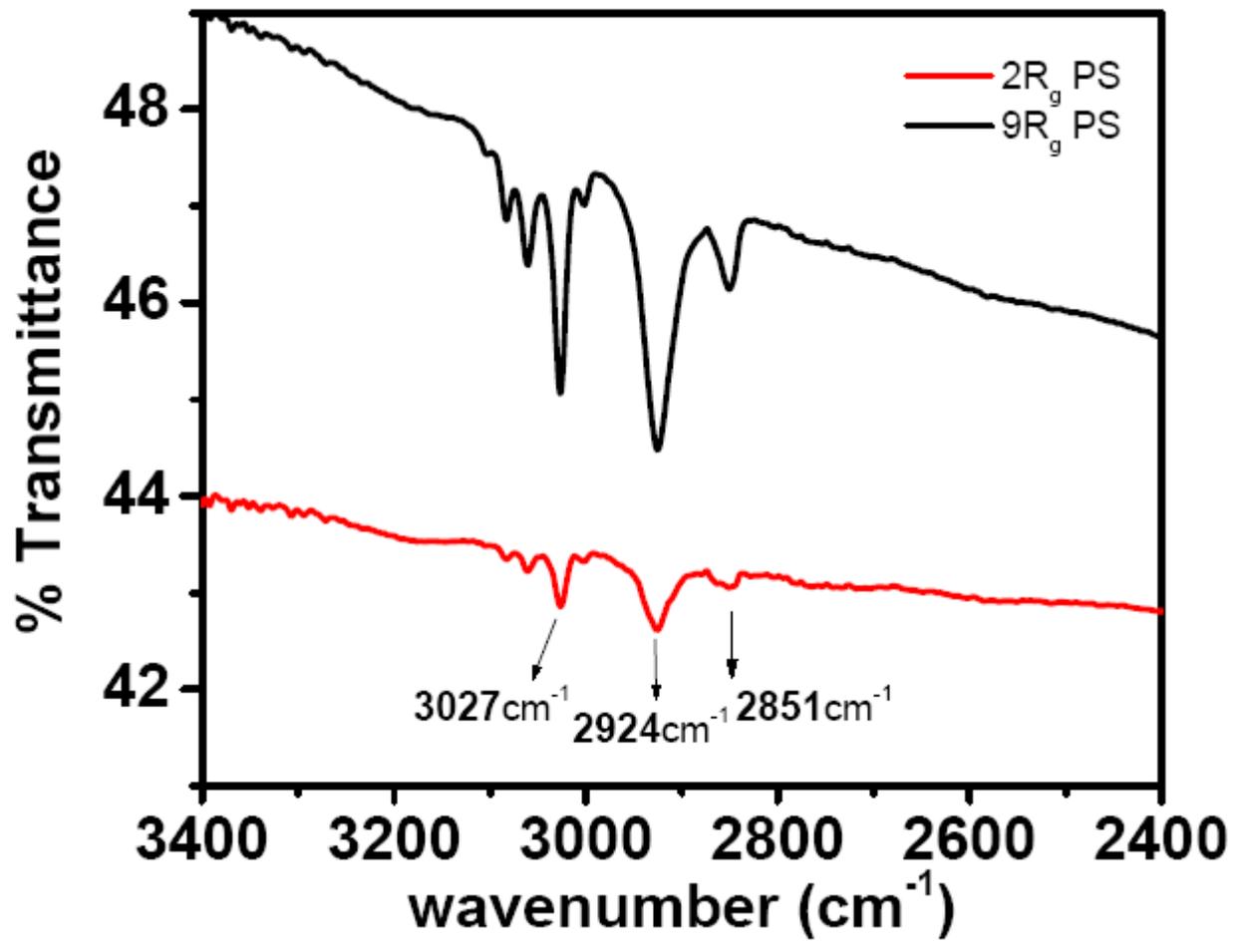

Figure 1

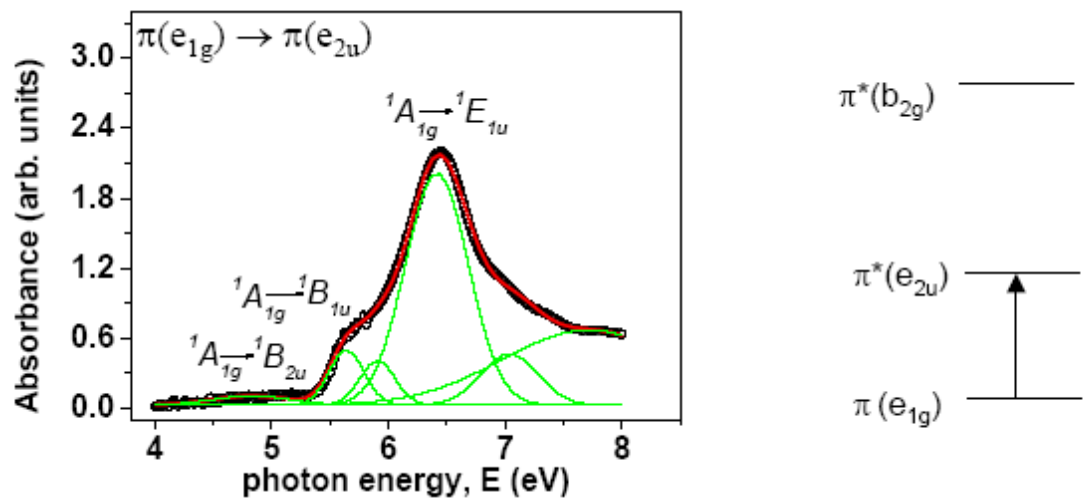

Figure2a

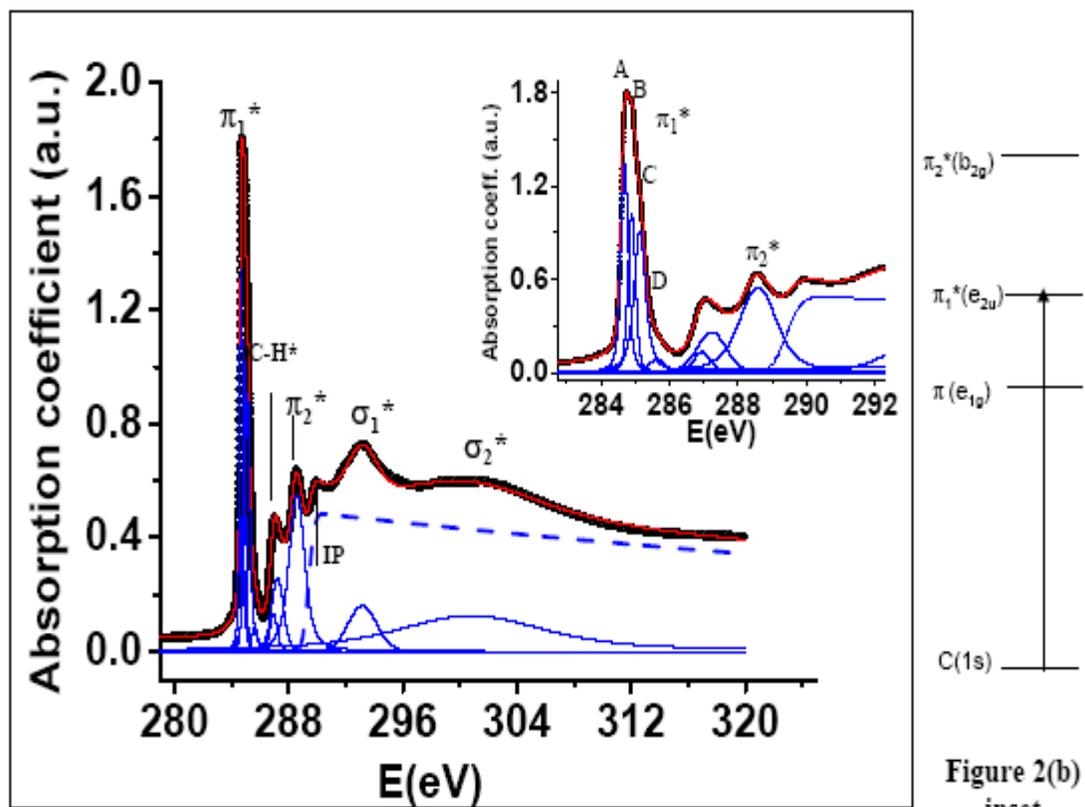

Figure2b

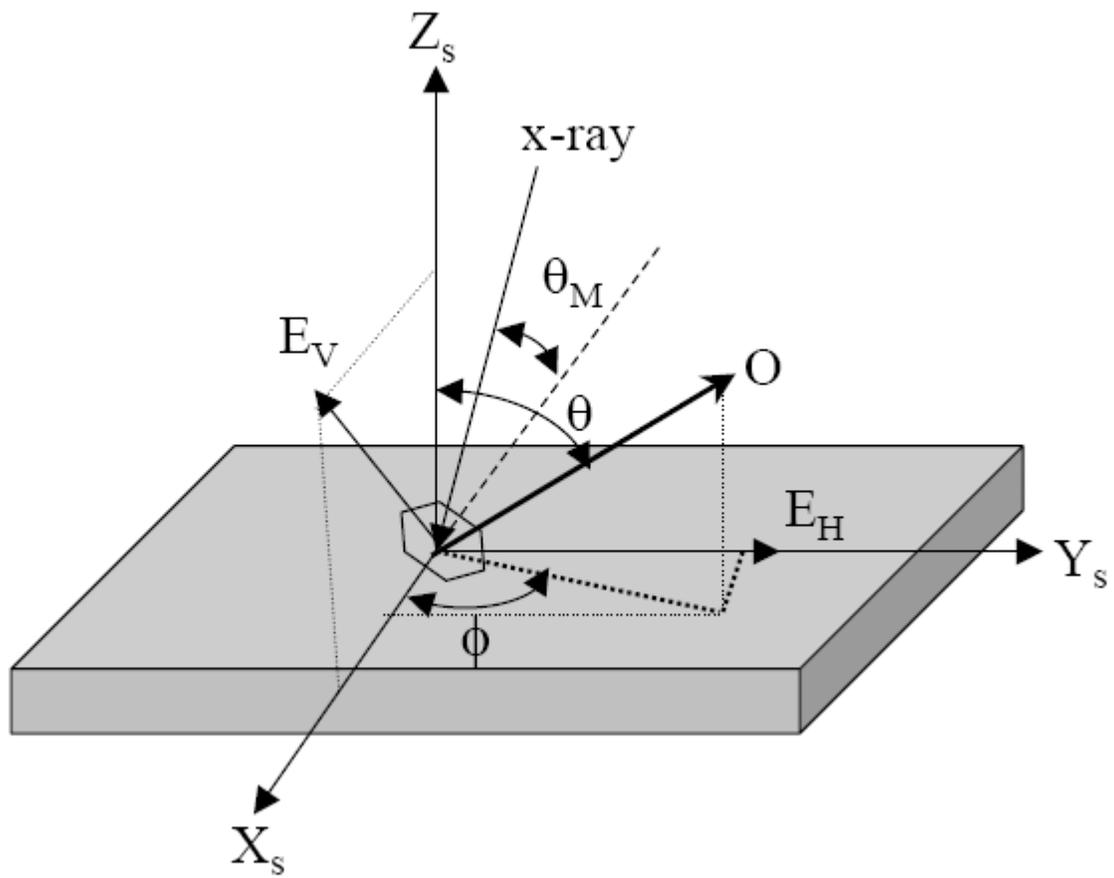

Figure 3a

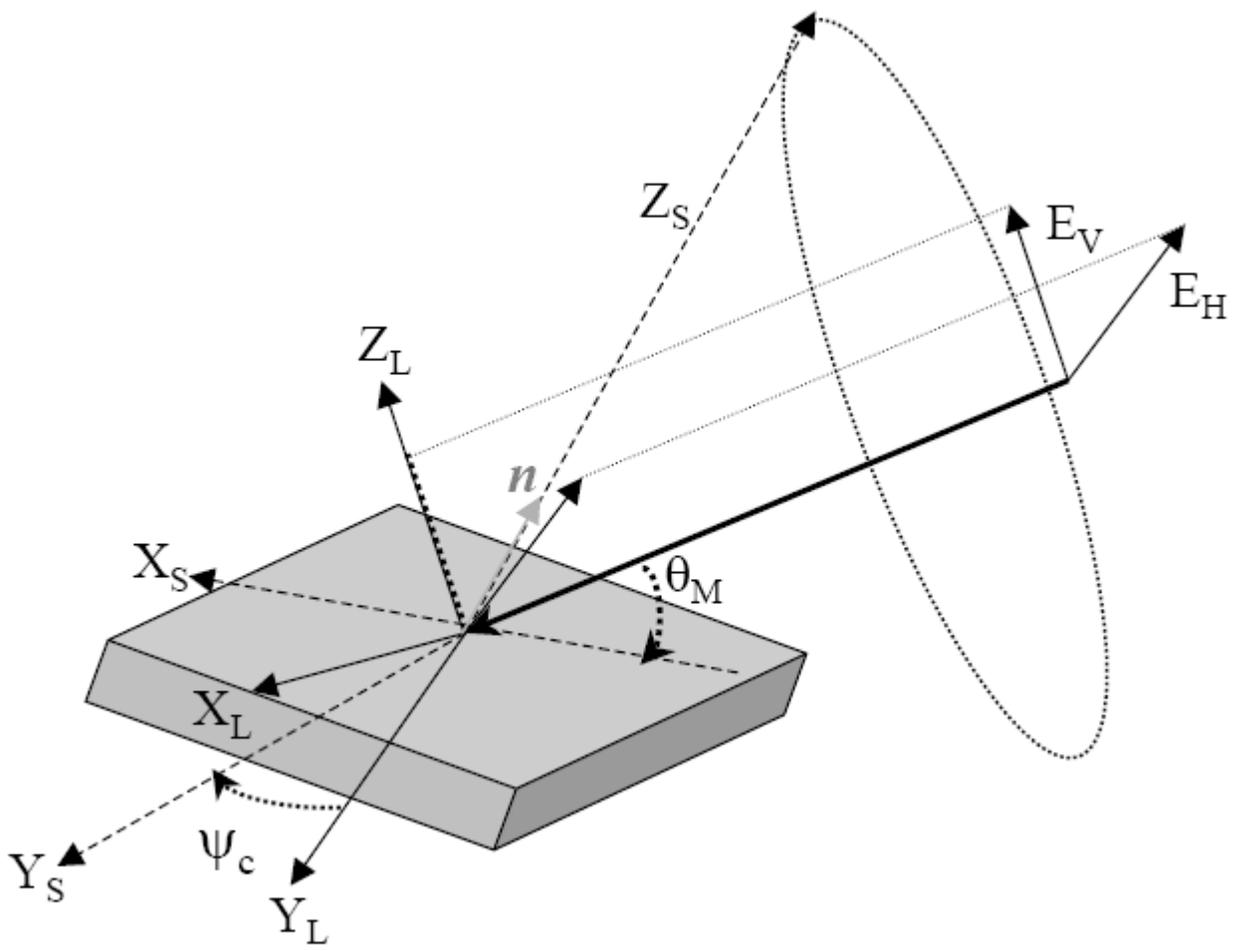

Figure 3b

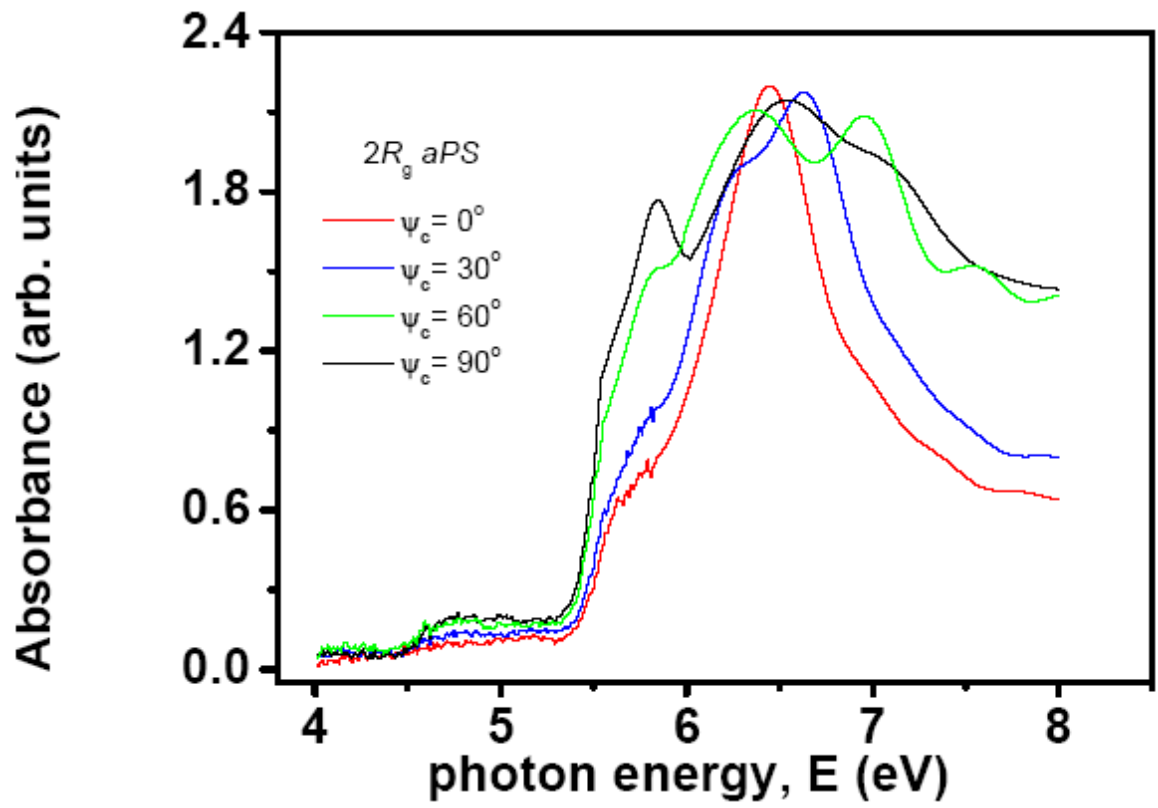

Figure4a

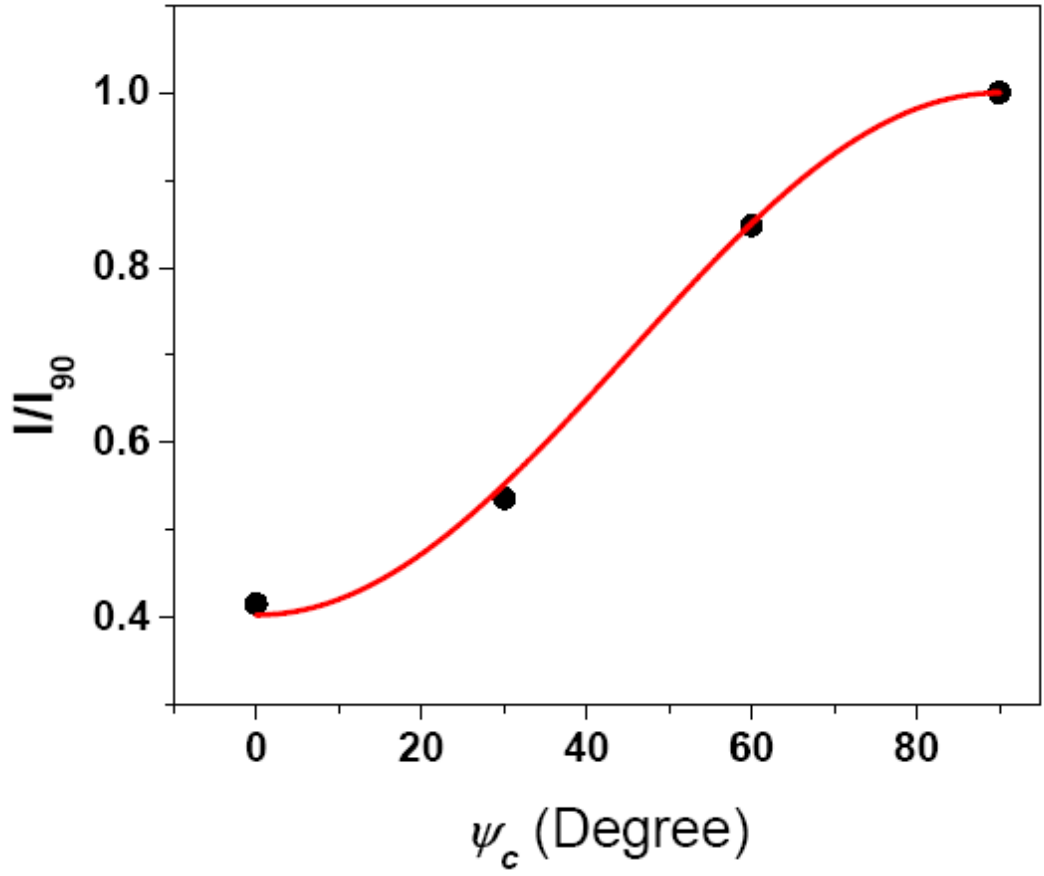

Figure4b

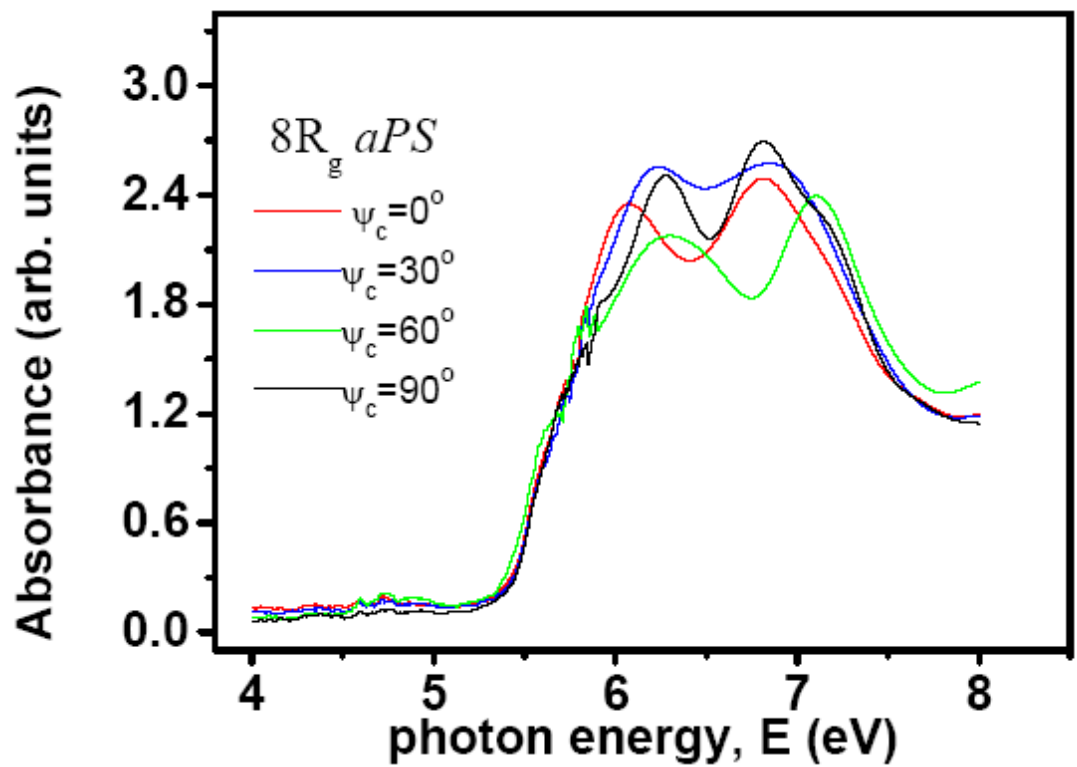

Figure4c

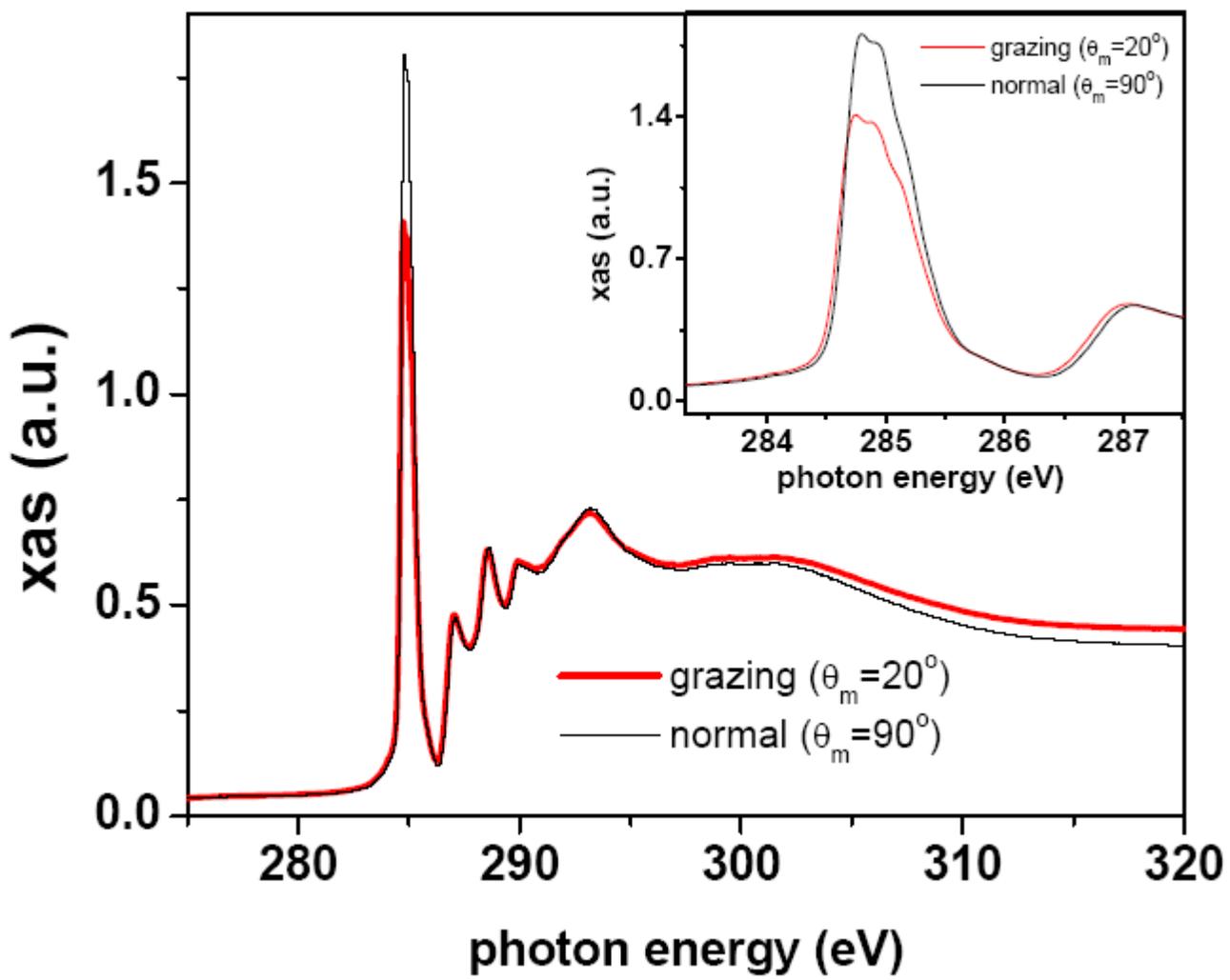

Figure 5a

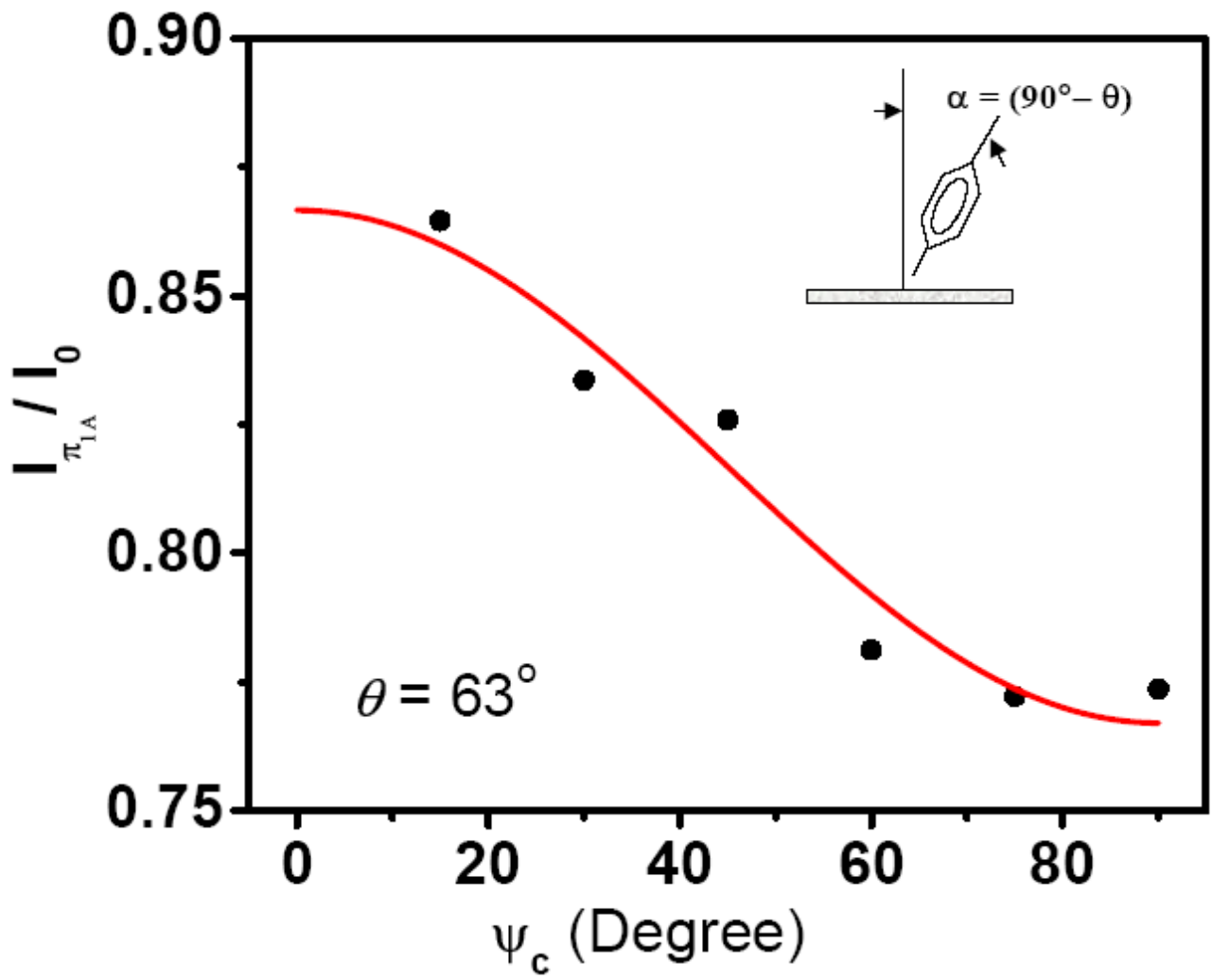

Figure5b